\documentstyle[aps,prd,preprint,floats,tighten,eqsecnum,epsf]{revtex}

\newcommand{\beq}{\begin{equation}}
\newcommand{\eeq}{\end{equation}}
\newcommand{\beqa}{\begin{eqnarray}}
\newcommand{\eeqa}{\end{eqnarray}}

\begin{document}
\draft

\preprint{\vtop{\hbox{RU00-1-B}\hbox{UM-TH-00-05}\hbox{hep-ph/0003199}}}

\title{Non-Fermi Liquid Behaviour, the BRST Identity in the
Dense Quark-Gluon Plasma and Color Superconductivity.}
\author{William E. Brown${}^a$, James T. Liu${}^{b}$
and Hai-cang Ren${}^{a,c}$}
\address{${}^a$ Department of Physics, The Rockefeller University,\\
 1230 York Avenue, New York, NY 10021.}
\address{${}^b$ Randall Laboratory of Physics, University of Michigan,\\
Ann Arbor, MI 48109.}
\address{${}^c$ Department of Natural Science, Baruch College of CUNY,\\
New York, NY 10010.}

\maketitle

\begin{abstract}
At sufficiently high baryon densities, the physics of a dense quark-gluon
plasma may be investigated through the tools of perturbative QCD.  This
approach has recently been successfully applied to the study of color
superconductivity, where the dominant di-quark pairing interaction arises
from one gluon exchange.  Screening in the plasma leads to novel behaviour,
including a remarkable non-BCS scaling of $T_C$, the transition temperature
to the color superconducting phase.  Radiative corrections to one gluon
exchange were previously considered and found to affect $T_C$.  In
particular, the quark self-energy in a plasma leads to non-Fermi liquid
behaviour and suppresses $T_C$.  However, at the same time, the quark-gluon
vertex was shown not to modify the result at leading order.  This dichotomy
between the effects of the radiative corrections at first appears rather
surprising, as the BRST identity connects the self-energy to the vertex
corrections.  Nevertheless, as we demonstrate, there is in fact no
contradiction with the BRST identity, at least to leading log order.  This
clarifies some of the previous statements on the importance of the higher
order corrections to the determination of $T_C$ and the zero temperature gap
in color superconductivity.

\end{abstract}

\pacs{PACS numbers: 12.38Aw, 12.38.-t, 11.10.Wx, 11.15.Ex}

\widetext

\section{Introduction.}

Attention to the physics of a dense quark-gluon plasma has recently
been revived, along with progress in understanding the phase structure of
QCD.  Interest has only heightened recently with the projected onset
of relativistic heavy ion collisions at BNL.  Such unusual conditions of
QCD may also exist in nature, for example in the core of a dense neutron
star. The asymptotic freedom of QCD makes a perturbative
treatment applicable at sufficiently high baryon density, and the
attractive di-quark interaction mediated by one gluon exchange
in the antisymmetric color representation induces superconductivity below
a certain temperature
\cite{bailin1984,alford1998a,rapp1998,alford1999,schafer1998,pr1,pr2}.

Working at non-zero temperature and chemical potential introduces several
complications.  One of the primary features of the plasma is that it screens
the QCD interaction.  Thus it is necessary in principle to dress gluon
propagators with hard dense/thermal loops (HDL/HTL) in the plasma.  For
conditions in the range of interest for color superconductivity, the
temperature effects are less important, and only the effects of screening
by HDL need to be considered.  That this screening is important is
a corollary of a more general statement, namely that, with a long range
interaction at non-zero chemical potential, a straightforward power series
expansion of the free energy in the coupling $g$ results in infra-red
divergences.  A resummation over the fermion loops, and the replacement
of the bare gluon propagator by that dressed with HDL
\cite{hdl1,hdl2,hdl3}, prior to
perturbative expansion resolves the infra-red difficulties.  This was
demonstrated, for example, for a non-relativistic electron gas with
Coulomb interaction in \cite{Gell-Mann}. The resultant perturbative series
contains logarithms of the coupling constant accompanying the powers of it.

As a result of HDL, the electric gluon propagator is screened
effectively by a Debye mass, $m_D$, while the magnetic propagator is
poorly screened via Landau damping in the sensitive region of momentum
space. An important consequence of this is the introduction of non-Fermi
liquid behavior of the quark self-energy \cite{son1998}.  In the infra-red
limit (highlighted by a cutoff $l_c$) this self-energy is
\beq
\label{eq:nfl}
\Sigma(\nu,\vec p\,)\vert_{p=\mu}\simeq-\frac{ig^2}{12\pi^2}
C_f\gamma_4\,\nu\ln\frac{4l_c^3}{\pi m_D^2|\nu|},
\eeq
where $(\vec p,\nu)$ are the Euclidean energy-momentum. Such a
non-analytic dependence on energy $\nu$ was first discovered in solid
state physics \cite{pincus} in the context of magnetic interactions.
The logarithm suppresses the
quasi-particle weight at the Fermi level and the single fermion
occupation number becomes a continuous function at $p=p_F$, the Fermi
momentum, in contrast to the kink of a Fermi liquid. Another effect is
a term $\sim T\ln T$ in the specific heat, but this turns out to be
too small to be observed since the magnetic coupling represents merely
a relativistic correction.  Such non-analyticity as indicated by
(\ref{eq:nfl}) was also suggested for certain strongly correlated
systems such as high $T_C$ superconductors \cite{varma}.

In a relativistic quark-gluon plasma, the relatively poor screening by
Landau damping is far more transparent. The di-quark pairing force is
dominated by magnetic gluons, and Landau damping gives rise to a
remarkable non-BCS scaling of the transition temperature and the
energy gap for color superconductivity
\cite{son1998,schafer1999,pisarski1999b,hong},
\beq
\label{eq:scl}
k_BT_C=c\frac{\mu}{g^5}e^{-\sqrt{\frac{6N_c}{N_c+1}}\frac{\pi^2}{g}}.
\eeq
The non-Fermi liquid behavior (\ref{eq:nfl}) suppresses the pre-exponential
factor $c$ significantly, and we found that \cite{BLR1}
\beq
\label{eq:ratio}
c=c_0e^{-\frac{1}{16}(\pi^2+4)(N_c-1)}\simeq 0.176\,c_0 \qquad\hbox{for
$N_c=3$},
\eeq
with $c_0$ the pre-exponential factor without radiative corrections
\cite{schafer1999,pisarski1999b,pisarski1999c,BLR2}.  We have also argued
that the contributions from other higher order diagrams to (\ref{eq:ratio})
are subleading.

Since the inverse quark propagator is related to the quark-gluon vertex
function through a BRST identity, so is the quark self-energy $\Sigma(P)$
of Fig.~\ref{fig1} to the radiative correction $\Lambda_\mu^l(P^\prime,P)$
of Fig.~\ref{fig2}.
To illustrate such a relation, we focus on the vertex correction in
Fig.~\ref{fig2}a,
which survives in the abelian case and will be referred to as the `abelian'
vertex in the following. After factorizing out the group theoretic
coefficients from the vertex and self-energy,
\beq
\label{factor1}
\Lambda_\mu^{l(a)}(P^\prime,P)=gT_f^mT_f^lT_f^m\Lambda_\mu(P^\prime,P)
\eeq
and
\beq
\label{factor2}
\Sigma(P)=T_f^lT_f^l\Xi(P),
\eeq
we have the Takahashi identity
\beq
\label{tata}
(P^\prime-P)_\mu\Lambda_\mu(P^\prime,P)=\Xi(P^\prime)-\Xi(P).
\eeq
Taking the limit $P^\prime\to P$, we end up with the usual Ward identity,
\beq
\label{eq:ward}
\Lambda_\mu(P,P)=\frac{\partial}{\partial P_\mu}\Xi(P),
\eeq
which is similar to the Ward identity of QED.
Because of the above behavior of $\Xi(P)$, this
identity raises a suspicion that $\Lambda_\mu(P^\prime,P)$ must also
contribute to the pre-exponential factor in a manner similar to
(\ref{eq:ratio}). This was, however, ruled out in \cite{BLR1} where we
showed that while the derivative $\frac{\partial}{\partial\nu}\Xi(P)$
contains the logarithm of $\nu$, the derivative
$\frac{\partial}{\partial p_i}\Xi(P)$ does not, and thus the effect is not
that of wave function renormalization. The contribution of
$\Lambda_4(P^\prime, P)$ remains subleading even with the logarithm since
the Coulomb propagator attached to it is strongly screened.

Nevertheless a paradox arises here. The integral
representations of $\Lambda_i(P,P)$ and $\Lambda_4(P,P)$ look identical while
the above results, together with (\ref{eq:ward}), suggests different answers.
In this article we shall disentangle this mystery. It turns out that the
expression $\Lambda_\mu(P,P)$ is highly ambiguous in the presence of a
Fermi sea, and in particular,
\beq
\label{eq:limit}
\lim_{\vec p\,^\prime\to\vec p}\lim_{\nu^\prime\to\nu}\Lambda_\mu(P^\prime,
P)\neq \lim_{\nu^\prime\to\nu}\lim_{\vec p\,^\prime\to\vec p}
\Lambda_\mu(P^\prime,P),
\eeq
a common ambiguity of the infra-red limit (the zero energy-momentum limit of
soft lines of a diagram) in the absence of covariance. The contribution of
$\Lambda_\mu^l(P^\prime,P)$ to color superconductivity comes mainly
in the region $|\nu^\prime-\nu|\ll\mu$ with $|\vec p\,^\prime-\vec p\,|\sim
m_D^{2/3}|\nu^\prime-\nu|^{1/3}$ while $|p-\mu|\sim|\nu|$ and
$|p^\prime-\mu|\sim|\nu^\prime|$, which is closer to the order of the left
hand side of (\ref{eq:limit}). By carefully tracing the subtleties of the
infra-red limit along the different routes, we are able to reconcile the
logarithmic behavior of (\ref{eq:nfl}) with the Ward identity (\ref{eq:ward})
as well as the full BRST identity when the group theoretic factors and the
vertex diagrams in Fig.~\ref{fig2}b and Fig.~\ref{fig2}c are restored. Yet
the suppression in (\ref{eq:ratio}) remains intact.

Though we are mainly addressing QCD and color superconductivity in
this article, the non-Fermi liquid
behavior of the fermion self-energy and the vertex function apply, to a
simpler extent, to the relativistic electron plasma as well. Such a plasma
exists
inside a white dwarf star, a supernova or a red giant star, for which the
condition that the chemical potential is much higher that the temperature
is valid.

In the next section, we shall calculate the quark self-energy and pin down
the mathematical mechanism behind the logarithm of (\ref{eq:nfl}). The vertex
function $\Lambda_\mu^l(P^\prime,P)$ is analyzed in section III in light of
the BRST identity. The contribution to color superconductivity will be
discussed in section IV together with some concluding remarks.

\section{The Quark Self-Energy.}

\begin{figure}[t]
\epsfxsize 4.5cm
\centerline{\epsffile{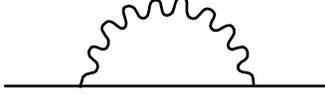}}
\bigskip
\caption{The quark self-energy diagram.}
\label{fig1}
\end{figure}
Little motivation is required for the analysis of the quark
self-energy, represented in Fig.~\ref{fig1}; it may appear as a simple
radiative correction in perturbative processes, but it also enters
the BRST identity.  The form of the self-energy also characterises
the non-Fermi liquid behaviour at high density and has a subtle
influence over the divergences of the quark vertices.  Without more ado,
in Euclidean space we write,
\begin{equation}
\Sigma(P) = -g^2T_f^l T_f^l \frac{1}{\beta} \sum_n \int
\frac{d^3\vec{l}}{(2\pi)^3} {\mathcal D}_{\mu\nu}(L)
\gamma_{\mu}S(L+P) \gamma_{\nu},
\end{equation}
where
$L = (\vec{l},-\omega_n)$, $P = (\vec{p},-\nu_n)$,
$\omega_n = n \epsilon$, $\nu_n = \left(n + \frac{1}{2}\right) \epsilon$,
$\epsilon = 2 \pi k_BT$ and $T_f^l T_f^l = C_f=\frac{N_c^2-1}{2N_c}$ for
$T_f$ in the fundamental representation of $SU(N_c)$.  Following the
notation of \cite{BLR1,BLR2}, we write the quark propagator as
\beq
S(P) = \frac{i}{/\kern-8pt P},
\eeq
where $/\kern-8pt P = \gamma_4(\mu+i\nu)-i\vec\gamma\cdot\vec p$.
In the presence of a Fermi sea it is necessary to incorporate HDL into
the gluon propagator at leading order.  While
it is possible that a magnetic mass of order $T$ exists, at high
density $\mu \gg k_BT$ the damping due to HDL prevails over that due
to HTL \cite{Bellac}.  Incorporating HDL, the gluon propagator in the
covariant gauge takes the form
\beq
\label{eq:glue2}
{\cal D}_{\mu\nu}(K)=\frac{-i}{K^2+\sigma^M(k,\omega)}
P_{\mu\nu}^T+\frac{-i}{K^2[1+
\sigma^E(k,\omega)/ k^2]}P_{\mu\nu}^L
-i\alpha\frac{K_\mu K_\nu}{(K^2)^2},
\eeq
where $K=(\vec k,-\omega)$, $K^2=k^2+\omega^2$, $P_{ij}^T=
\delta_{ij}-\hat k_i\hat k_j$, $P_{i4}^T=P_{4j}^T=P_{44}^T=0$,
\beq
P_{\mu\nu}^L=\delta_{\mu\nu}-\frac{K_\mu K_\nu}{K^2}-P_{\mu\nu}^T,
\eeq
and $\alpha$ is the gauge parameter (we have adopted the notation $|\vec
k\,|=k$). The electric self-energy $\sigma^E( k,\omega)$ and the
magnetic self-energy $\sigma^M(k,\omega)$ in (\ref{eq:glue2}) are given
by $\sigma^E(k,\omega) = m_D^2
f^E\left(\omega/k\right)$, $\sigma^M(k,\omega) =
m_D^2 f^M\left(\omega/k\right)$, with $m_D^2 \simeq \frac{N_f
g^2 \mu^2}{2 \pi^2}$ and
\begin{eqnarray}
\label{eq:F}
f^E(x) &=&   \left[ 1 - x \tan^{-1} \left( \frac{1}{x}
\right) \right],\\
\label{eq:G}
f^M(x) &=& \frac{x}{2} \left[ (1+x^2) \tan^{-1} \left(
\frac{1}{x} \right) - x \right].
\end{eqnarray}
For more discussion of our notation or HDL in general see \cite{BLR1,BLR2}
or \cite{Bellac}, respectively.  From (\ref{eq:F}) we can see that
the Coulomb interaction is strongly screened while from (\ref{eq:G})
the magnetic interaction is not.  In this paper, we shall be
interested only in the leading infra-red behavior, which comes solely
from magnetic gluon exchange, and so we shall neglect the electric
contributions and regard
\beq
{\cal D}_{\mu\nu}(K)\approx-i{\cal D}(k,\omega)P_{\mu\nu}^T,
\eeq
with
\beq
\label{eq:calddef}
{\cal D}(k,\omega)=\frac{1}{k^2+\omega^2+\sigma^M(k,\omega)}.
\eeq

To focus upon the infra-red behavior, we separate the loop integral into two
regions and rewrite the self-energy as
\beq
\Sigma(P) = C_f \left[ \Xi^<(P) + \Xi^>(P) \right],
\eeq
where the superscripts denote integration inside and outside the
infra-red sensitive region: $0<l<l_c$, $-\omega_c < \omega_n <
\omega_c$ with $l_c,\omega_c \ll \mu$.  We shall evaluate the infra-red
sensitive region only,
\beq
\Xi^<(P) \simeq -\frac{g^2}{4\pi^2} \int_0^{l_c} dl\,
l^2 \int_{-1}^{1} d(\cos \theta) \frac{1}{\beta} \sum_n \frac{\gamma_4
-i(\hat{l}\cdot\hat{p})^2\vec\gamma\cdot\hat p}{\xi - i(\omega_n + \nu_m)}
{\cal D}(l,\omega_n),
\eeq
where $\xi = |\vec l + \vec p\,| - \mu$.
Corrections due to the change from a discrete sum to an integral are
sub-leading; they can be obtained using zeta-function techniques as
demonstrated for similar processes in \cite{BLR2}.  Thus, in pursuit of
the leading order behavior only, we immediately move to the continuous
energy limit with $k_BT\ll\nu_m\ll\omega_c$.  Making the change of variables
from $\theta$ to $\xi$,
\beq
\label{cov}
\int_{-1}^{1} d(\cos \theta) [\gamma_4 - i(\hat{l}\cdot\hat{p})^2
\vec\gamma\cdot\hat p]
\simeq \int_{\xi_p - l}^{\xi_p + l} \frac{d\xi}{l}
\left[ \gamma_4 - i\frac{\mu^2}{l^2 p^2} \left(\xi - \xi_p \right)^2
\vec\gamma\cdot\hat p\right],
\eeq
with $\xi_p = p - \mu$, we find,
\beqa
\label{212}
& &\Xi^<(\nu,\vec p\,) \simeq -\frac{g^2}{8\pi^3} \int_0^{l_c} dl\,l
\int_{-\omega_c}^{\omega_c} d\omega {\cal D}(l,\omega) F(\nu,p;l,
\omega), \\
\label{213}
& &F(\nu,p;l,\omega) = \int_{\xi_p - l}^{\xi_p + l} d\xi
\frac{\gamma_4 - i\frac{\mu^2}{l^2 p^2} \left( \xi - \xi_p\right)^2
\vec\gamma\cdot\hat p}
{\xi - i (\omega + \nu)}.
\eeqa
Fixing $p = \mu$ for the external lines and carrying out the integration over
$\xi$, we have,
\beq
\label{eq:eff}
F(\nu,\mu;l,\omega)=2i\gamma_4\tan^{-1}\frac{l}{\omega+\nu}
+2\frac{\omega+\nu}{l}\vec\gamma\cdot\hat p\left(
1-{\omega+\nu\over l}\tan^{-1}\frac{l}{\omega+\nu}\right),
\eeq
so that
\beqa
\label{delt}
\frac{\partial}{\partial \nu} F(\nu,\mu;l,\omega) &=& 2 \pi i \gamma_4
\delta(\nu + \omega) - \frac{2il}{(\omega+\nu)^2+l^2}\gamma_4 \nonumber \\
&&\qquad+\frac{2}{l}\left[
-2\frac{\omega+\nu}{l}\tan^{-1}\frac{l}{\omega+\nu}+\frac{2(\omega+\nu)^2
+l^2}{(\omega+\nu)^2+l^2}\right]\vec\gamma\cdot\hat p,
\eeqa
where the delta function comes from the discontinuity of the inverse tangent
function. We find the energy
dependence of the self-energy by differentiating,
\beq
\left.\frac{\partial}{\partial \nu} \Xi^<(\nu,\vec p\,)\right|_{p=\mu}
= g^2[A(\nu)+B(\nu)],
\eeq
with
\beq
\label{215}
A(\nu) = -\frac{i}{4 \pi^2}\gamma_4 \int_0^{l_c} dl
\frac{l}{l^2 + \nu^2 + m_D^2f^M \left( \frac{-\nu}{l}\right)}
\eeq
and
\begin{eqnarray}
\label{B}
B(\nu) &=& \frac{1}{4\pi^3}\int_0^{l_c}dl\int_{-\omega_c}^{\omega_c}d\omega
{\cal D}(l,\omega)\\
&&\qquad\qquad\Biggl\{\frac{il}{(\omega+\nu)^2+l^2}\gamma_4
-\frac{1}{l} \left[2\frac{\omega+\nu}{l}\tan^{-1}\frac{l}{\omega+\nu}
+\frac{2(\omega+\nu)^2+l^2}{(\omega+\nu)^2+l^2}\right]\vec\gamma\cdot\hat p
\Biggr\}.\nonumber
\end{eqnarray}
Noting the asymptotic behavior, that $f^M(x) \simeq {\pi\over4}|x|$ for
$|x| \ll 1$, a scale $l_0$ may be introduced to divide the integration in
$A(\nu)$ into two: $|\nu| \ll l_0 \ll \left(m_D^2 |\nu|
\right)^{1/3}$.  For $l < l_0$ we have the contribution,
\beq
\int_0^{l_0} dl \frac{l}{l^2 + \nu^2 + m_D^2 f^M\left( \frac{\nu}{l} \right)}
\leq \frac{1}{m_D^2}\int_0^{l_0} dl \frac{l}{f^M\left( \frac{\nu}{l} \right)}
<\frac{1}{m_D^2 f^M\left( \frac{\nu}{l_0} \right)} \int_0^{l_0} dl\,l
\sim \frac{l_0^3}{m_D^2|\nu|} \ll 1.
\eeq
All of the inequalities follow straightforwardly from the definition
of $l_0$ except for the second, which is due to $f^M(\nu/l)$ being a
monotonically decreasing function of $l$.  Therefore, neglecting this
subleading contribution, we find a logarithmic infra-red singularity
in $A(\nu)$ arising from the second region (namely the region $l_0<l<l_c$).
The integration $B(\nu)$ is finite in the limit
$\nu\to 0$. We end up with
\beq
\left. \frac{\partial}{\partial \nu} \Xi^<(\nu,\vec p\,)\right|_{p=\mu}
 = -\frac{ig^2}{4 \pi^2}\gamma_4 \int_{l_0}^{l_c} dl
\frac{l^2}{l^3 + \nu^2 l + {\pi\over4}m_D^2|\nu|}\simeq -\frac{ig^2}{12 \pi^2}
\gamma_4\ln \frac{4l_c^3}{\pi m_D^2 |\nu|}+\cdots.
\eeq

That the self-energy does not depend upon the spatial momentum in the
infra-red limit can easily be ascertained by differentiating
$\Xi(P)$ with respect to $p_i$.  Noting that
$\frac{\partial}{\partial p_i} = \hat{p}_i \frac{\partial}{\partial
\xi_p}$, from (\ref{212}) and (\ref{213}) it is straightforward to find,
\beq
\left.\frac{\partial}{\partial \xi_p} \Xi^<(\nu,\vec{p}\,) \right|_{p=\mu} =
ig^2B(\nu),
\eeq
which is both real and finite in the limit $\nu \to 0$. Therefore the
logarithmic singularity can not be attributed to a wavefunction
renormalization. This is also the case found in a solid state physics
context \cite{pincus}.

To summarise, we find that in a dense quark-gluon plasma the quark
self-energy exhibits non-analytic behavior only for the energy component,
\beq
\label{self}
\left.\Sigma(P)\right|_{p=\mu} = -\frac{ig^2}{12\pi^2}
C_f \gamma_4\,\nu \ln \frac{4l_c^3}{\pi m_D^2 |\nu|} + \cdots
\eeq
(the imaginary part of the self-energy, contributing to damping in the
plasma \cite{damp1,damp2}, is analytic as $\nu\to0$).
It is important to note that the infra-red non-analyticity in the
energy originates in the discontinuity of the pole cutting the contour
in the $\xi$ integration.  This feature gives rise to the
$\delta$-function in (\ref{delt}) which ultimately leads to the
infra-red non-analyticity.  This behavior will be seen to repeat itself
in section IIIA where it will lead to infra-red divergences in
the radiative corrections to the quark-gluon vertex.

{}From the result (\ref{self}) it is clear that covariance is broken;
this is a direct effect of the presence of a Fermi sea.  We
also see that $\Xi^<(0) = 0$, so that the self-energy leads to
no chemical potential renormalisation from the infra-red side.
What is also of considerable interest is the BRST identity.  How this is
met in the dense quark-gluon plasma is subtle and we investigate this
phenomenon in the next section.
%

\section{The BRST Identity at High Density.}

Since the quark self-energy is only non-analytic in the external
energy, one may expect from a generalisation of the Ward
identity of QED that the Coulomb-quark vertex has similar behavior and
is divergent while the magnetic gluon-quark vertex is not.  However,
it is also clear that the integrands in the infra-red region for
$\Lambda_4^l$ and
$\Lambda_i^l$ are mathematically identical, save for an indiced
prefactor.  Apparently we have something of a paradox: from the
self-energy we expect only the Coulomb vertex to be divergent, but
there appears to be no mathematical difference between the infra-red
contribution to the Coulomb and magnetic vertices.  We shall resolve
this paradox in this section and show how
the BRST identity works at high density.  First of all, we shall examine the
zero energy-momentum transfer limit of the abelian vertex
$\Lambda_\mu(P^\prime,P)$ and derive the precise expression of the Ward
identity (\ref{eq:ward}):
\beqa
\label{Ward}
\lim_{\nu' \to \nu} \lim_{\vec{p}\,' \to \vec{p}} \Lambda_4(P',P) &=&
-\frac{\partial}{\partial \nu} \Xi(P), \\ \nonumber
\lim_{\vec{p}\,' \to \vec{p}} \lim_{\nu' \to \nu} \Lambda_i(P',P) &=&
\frac{\partial}{\partial p_i} \Xi(P).
\eeqa
The latter relation for the magnetic vertex was previously investigated
in \cite{Chakra} for a system of fermions interacting with transverse
abelian gauge bosons.  Contrasting the Coulomb and magnetic cases in
(\ref{Ward}) provides an important clue hinting that the ordering of
limits contributes to the subtlety we are overlooking.  
To show how the paradox is resolved this identity
shall be considered in the next subsection, where the infra-red behavior of
the vertices is discussed and the abelian vertex is treated in detail.
In the second subsection we shall show how the full BRST identity
works in the dense quark-gluon plasma in terms of Feynman
diagrams.

\subsection{Infra-red Behavior of the Abelian Vertex.}

To explore the behavior of the quark-gluon vertices and their relation
to the BRST identity we shall analyse in detail the abelian vertex
$\Lambda_{\mu}^{l(a)}$ shown in Fig.~\ref{fig2}a.  We refer to this vertex
as `abelian' since it is the only physical vertex that also appears in the
abelian theory.  In order to simplify matters further, in this
subsection we shall put both external quarks on-shell, $p=p'=\mu$.

Intuitively, we may expect the behavior of the vertex to depend subtly
upon the ordering of the limits.  We may develop this intuition from
HDL, for example, where although there is an analytic result for the
screening
(\ref{eq:F}) and (\ref{eq:G}), it has different asymptotic behavior in
the two orderings of the limits ($x\to 0$ and $x \to \infty$).  As we
shall see, HDL and the BRST identity at high density are intimately
connected and it is no surprise that the ordering of limits in
(\ref{Ward}) is crucial in resolving the paradox.  We write the
abelian vertex as
\begin{eqnarray}
\Lambda_\mu^{l(a)}(P',P)&=&gT_f^mT_f^lT_f^m\Lambda_\mu(P',P)\nonumber\\
&=&gT_f^l\left(-{C_{ad}\over2}+C_f\right)\Lambda_\mu^{(a)}(P',P),
\end{eqnarray}
where
\beq
\label{abelian}
\Lambda_{\mu}^{(a)}(P',P) = \frac{ig^2}{\beta} \sum_n \int
\frac{d^3 \vec l}{(2 \pi)^3} {\cal D}_{\nu \rho}(l,\omega)
\gamma_{\nu} S(L+P') \gamma_{\mu}  S(L+P) \gamma_{\rho}.
\eeq
This may be written in terms of two integrals, one
inside and one outside the infra-red sensitive region; $0<l<l_c$,
$-\omega_c < \omega < \omega_c$ with $l_c,\omega_c \ll \mu$,
\beq
\Lambda_{\mu}^{(a)}(P',P) =
\hat P_\mu \left[ \Lambda^{(a)<}(P',P) + \Lambda^{(a)>}(P',P) \right].
\eeq
with $\hat P_\mu=(-i\hat p, 1)$.
In this subsection we are only interested in the leading infra-red
behavior. So we evaluate,
\beqa
\label{lambda1}
\Lambda^{(a)<}(P',P) &\simeq& \frac{g^2}{8\pi^3} \int_0^{l_c} dl\,l^2
\int_{-1}^{1} d (\cos \theta)
(\gamma_4-i\vec\gamma\cdot\hat p\cos^2\theta)
 \widetilde{\Lambda}(P,P';L), \\
\label{lambdatilde}
\widetilde{\Lambda}(P,P';L) &=& \int \kern-10pt \circ
\frac{d\omega}{2\pi} \frac{{\cal D}(l,\omega)}{\zeta' - \zeta}
\left(\frac{1}{\omega + \zeta'} - \frac{1}{\omega - \zeta'}
- \frac{1}{\omega + \zeta} + \frac{1}{\omega - \zeta}
\right)\ln (-\omega),
\eeqa
where $\zeta = \nu + i \xi$, $\xi = |\vec{l} + \vec{p}\,| - \mu$ and
$\zeta'$ and $\xi'$ refer to $\nu'$, $\vec{p}\,'$.  The logarithm in
(\ref{lambdatilde}) introduces a branch cut which we may take to lie
along the positive real axis and the contour to run above and below it
in the normal fashion.  As shown for the self-energy, it is only
the discontinuities that occur as poles cut the contour and branch cut
that induce the infra-red singularity. Hence we need only focus
upon the second and fourth terms in (\ref{lambdatilde}), since the other
terms are regular.  Using the convention that ${\rm arg}(-1) = 0$, we find
that the contribution of these poles reads
\beq
\widetilde{\Lambda}_{\rm{disc.}}(P',P;L) = \frac{\pi}{\zeta' - \zeta}
\left[ {\rm Sign}(\xi) {\cal D}(l,\zeta)
- {\rm Sign}(\xi') {\cal D}(l,\zeta') \right],
\eeq
where the sign function comes from the discontinuity of $\ln(-\omega)$
crossing the cut.

We are now in a position to take the limit $P_{\mu} \to P_{\mu}'$ for
the Ward identity and we shall consider the two different orderings of
the limits in turn:
\beqa
{i)} ~~\lim_{\nu' \to \nu} \lim_{\vec{p}\,' \to \vec{p}}
\widetilde{\Lambda}_{\rm{disc.}}(P', P;L) &=& - \pi {\rm Sign}(\xi)
\frac{\partial}{\partial \nu} {\cal D}(l,\zeta), \\
{ii)} ~~\lim_{\vec{p}\,' \to \vec{p}} \lim_{\nu' \to \nu}
\widetilde{\Lambda}_{\rm{disc.}}(P',P;L) &=& i\pi \frac{\partial}{\partial \xi}
\left[ {\rm Sign}(\xi) {\cal D}(l,\zeta)\right].
\eeqa
In both cases we are looking at the infra-red limit, and thus fix the
external momentum to be $p = p' = \mu$, $\xi_p = 0$.

First of all, considering case $i$), using the change of variables
(\ref{cov}) it is straightforward to find,
\beqa
\left.\lim_{\nu' \to \nu} \lim_{\vec{p}\,' \to \vec{p}}
\Lambda^{(a)<}(P',P) \right|_{p = \mu} &=& -\frac{g^2}{8\pi^2}
\int_0^{l_c} dl\,l
\int_{-l}^{l} d\xi\>{\rm Sign}(\xi) \frac{\partial}{\partial \nu}
{\cal D}(l , \zeta)(\gamma_4-i\frac{\xi^2}{l^2}\vec\gamma\cdot\hat p) \\
\label{logs}
&=& \frac{ig^2}{4\pi^2}\gamma_4 \int_0^{l_c} dl\,l \left[ {\cal D}(l,\nu) -
{\cal D}(l,\nu+il) \right] + \cdots \ \\
\label{ans}
&=& \frac{ig^2}{12 \pi^2}\gamma_4 \ln \frac{4l_c^3}{\pi m_D^2 |\nu|} + \cdots,
\eeqa
where the second term in the brackets of (\ref{logs}) contributes to the
subleading terms denoted by ellipses in (\ref{ans}).

Secondly, considering case $ii$), we find that differentiation gives two
terms which will cancel in the leading order,
\begin{eqnarray}
\left. \lim_{\vec{p}\,' \to \vec{p}} \lim_{\nu' \to \nu}
\Lambda^{(a)<}(P',P)\right|_{p = \mu} = \frac{ig^2}{8\pi^2}\int_0^{l_c} dl \,l
\int_{-l}^l d\xi &&
(\gamma_4-i{\xi^2\over l^2}\vec\gamma\cdot\hat p)\nonumber\\
&&\times\left[{\rm Sign}(\xi) \frac{\partial}{\partial \xi}
{\cal D}(l,\zeta) + 2 \delta(\xi) {\cal D}(l,\zeta) \right].
\end{eqnarray}
The first term is identical to that evaluated for case $i$) above.  With the
same approximation, the second term is,
\beq
-\frac{ig^2}{4\pi^2} \int_0^{l_c} dl \frac{l^2}{l^3 + {\pi\over4}m_D^2 |\nu|}
\simeq -\frac{ig^2}{12\pi^2} \ln \frac{4l_c^3}{\pi m_D^2 |\nu|}.
\eeq
The two leading contributions cancel and in this ordering of limits
the vertex is finite.

Now we can see how the paradox is resolved.  The spatial abelian
vertex considered with the ordering of the limits in case $ii$) is
finite, in agreement with the second part of the identity
(\ref{Ward}).

\subsection{The BRST Identity.}

\begin{figure}[t]
\epsfxsize 8cm
\centerline{\epsffile{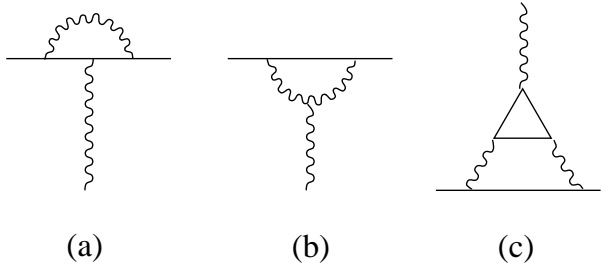}}
\bigskip
\caption{The physical radiative corrections to the quark-gluon vertex;
$a$) $\Lambda_{\mu}^{l(a)}$, the abelian vertex, $b$) $\Lambda_{\mu}^{l(b)}$,
the tri-gluon vertex and $c$) $\Lambda_{\mu}^{l(c)}$, the triangular vertex.}
\label{fig2}
\end{figure}

\begin{figure}[t]
\epsfxsize 8cm
\centerline{\epsffile{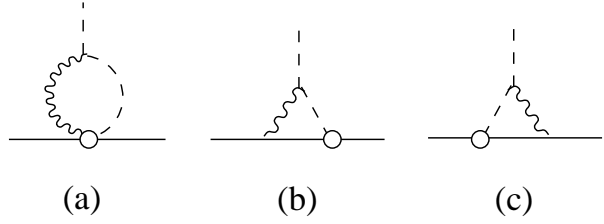}}
\bigskip
\caption{The non-physical ghost diagrams generated by the BRST
transformations; representing $a$) $R^{l(a)}$, $b$) $R^{l(b)}$ and $c$)
$R^{l(c)}$.  The open circles denote non-physical vertices
generated by BRST.}
\label{fig3}
\end{figure}

The BRST identity is a generalisation of the Ward-Takahashi identities
for non-abelian gauge theory obtained through the BRST
transformations.  The BRST version of the Takahashi identity can be written as,
\beq
\label{BRST}
(P' - P)^{\mu} \Lambda_{\mu}^l(P',P) = gT_f^l ( \Sigma(P') - \Sigma(P) )
+ R^l(P',P).
\eeq
The physical quark-gluon vertices $\Lambda_{\mu}^l =
\Lambda_{\mu}^{l(a)} + \Lambda_{\mu}^{l(b)} + \Lambda_{\mu}^{l(c)}$
are represented in Fig.~\ref{fig2}.  The non-physical ghost-quark vertices
induced by the BRST transformation, $R^l = R^{l(a)} + R^{l(b)} +
R^{l(c)}$, are represented in Fig.~\ref{fig3}. They vanish for on-shell
Minkowski momenta $P$ and $P^\prime$ at $\mu=0$.

The nontrivial part of the BRST identity (\ref{BRST}) is in the dressing
of the gluon lines of Figs.~\ref{fig1}, \ref{fig2} and \ref{fig3} by HDL
and the inclusion of Fig.~\ref{fig2}c.
The order of the perturbative expansion is mixed up without offsetting the
simple form of the identity. The detailed derivation of (\ref{BRST}) is
given in Appendix A.

Setting $\vec p\,^\prime=\vec p$ on the Fermi level, the BRST identity
(\ref{BRST}) implies that
\beqa
\label{eq:BRS4}
\lim_{\nu^\prime\to\nu}\left[\Lambda_4^l(P^\prime, P)-\frac{R^l(P^\prime,P)}
{\nu^\prime-\nu}\right]_{\vec p\,^\prime=\vec p,\; p=\mu}
&=&-gT_f^l\frac{\partial}{\partial\nu}\Sigma(P)\nonumber\\
&=&\frac{ig^3}{12\pi^2}C_f^{\vphantom{l}}T_f^l
\gamma_4\ln\frac{4l_c^3}{\pi m_D^2|\nu|}.
\eeqa
It follows from the discussions of the previous subsection that
$\Lambda_4^l(P^\prime,P)=\Lambda_4^{l(a)}(P^\prime,P)
+\Lambda_4^{l(b)}(P^\prime,P)
+\Lambda_4^{l(c)}(P^\prime,P)$ with
\beq
\label{ab4}
\Lambda_4^{l(a)}(P',P) = gT_f^l \left( -\frac{C_{ad}}{2} + C_f \right)
\frac{ig^2}{12\pi^2}\gamma_4 \ln \frac{4l_c^3}{\pi m_D^2|\nu|}.
\eeq
It remains to find the logarithmic terms from Figs.~\ref{fig2}b,
\ref{fig2}c or \ref{fig3} to reconcile the BRST identity (\ref{BRST})
to the leading order of the infrared logarithms.

For $\vec p\,^\prime=\vec p$ and $p=\mu$, we have
\beqa
\Lambda_4^{l(b)}(P^\prime,P) &=& ig^3f^{alb}T_f^aT_f^b
\frac{1}{\beta}\sum_n\int\frac{d^3\vec l}
{(2\pi)^3}(2\omega_n-\nu^\prime-\nu){\cal D}^2(l,\omega_n)
(\delta_{ij}-\hat l_i\hat l_j)\gamma_iS(P+L)\gamma_j \nonumber \\
&=& \frac{1}{2}gC_{ad}T_f^l[\Lambda_4^{(b)<}(P^\prime,P)
+\Lambda_4^{(b)>}(P^\prime,P)],
\eeqa
where the superscripts specify the contributions from loop momentum
inside and outside the infrared region, $|\omega|<\omega_c$ and
$l<l_c$. The inside contribution can be approximated by
\beqa
\Lambda_4^{(b)<}(P^\prime,P)&=&
\frac{g^2}{4\pi^3}\int_0^{l_c}dl\,l\int_{-\omega_c}^{\omega_c}d\omega
\,\omega{\cal D}^2(l,\omega)\int_{-l}^ld\xi\frac{\gamma_4-i\frac{\xi^2}{l^2}
\vec\gamma\cdot\hat p}{i(\omega+\nu)-\xi} \nonumber \\
&=& -\frac{g^2}{2\pi^3}(iI_i\gamma_4+I_2\vec\gamma\cdot\hat p),
\eeqa
where
\beq
I_1=\int_0^{l_c}dl\,l\int_{-\omega_c}^{\omega_c}d\omega\,
\omega{\cal D}^2(l,\omega)\tan^{-1}\frac{l}{\omega+\nu}
\eeq
and
\beq
I_2=\int_0^{l_c}dl\,l\int_{-\omega_c}^{\omega_c}d\omega\,
\omega{\cal D}^2(l,\omega)\left[\frac{\omega+\nu}{l}-
\tan^{-1}\frac{l}{\omega+\nu}\right].
\eeq
In the limit $\nu\to 0$, the integrand of both integrals $I_1$ and $I_2$
are positive and can be bounded by letting $\omega_c\to\infty$ and changing
the integration variables from $(l,\omega)$ to $(l,x=\omega/l)$. We have
\beq
|I_1|\leq\int_0^\infty dx
\frac{x\tan^{-1}{1\over x}}{(1+x^2)^2}\left[\ln\frac{l_c^2(1+x^2)}
{m_D^2f^M(x)}-1\right]
\eeq
and
\beq
|I_2|\leq\int_0^\infty dx
\frac{x\left(x-\tan^{-1}{1\over x}\right)}{(1+x^2)^2}
\left[\ln\frac{l_c^2(1+x^2)} {m_D^2f^M(x)}-1\right].
\eeq
Both integrals are convergent and hence $\Lambda_4^{l(b)}$ does not contribute
to the infrared logarithm. It is also straightforward to verify that the BRST
generated diagrams in Fig.~\ref{fig3} do not display any logarithmic
behavior in the limit $\nu\to 0$. Thus the only candidate left over is the
diagram in Fig.~\ref{fig2}c corresponding to gluon insertion on a HDL.

Though formidable as it looks, evaluation of Fig.~\ref{fig2}c can be
simplified with
the aid of a Ward type identity which relates the derivative of the gluon
self-energy and the tri-gluon vertex with three external gluon lines. Again the
answer is sensitive to the relative order of the limits $\nu^\prime\to\nu$ and
$\vec p\,^\prime\to \vec p$. In appendix B, we shall demonstrate that
\beq
\lim_{\nu^\prime\to\nu}\lim_{\vec p\,^\prime\to\vec p}
\Lambda_4^{l(c)}(P^\prime,P)=\frac{ig^3}{24\pi^2}C_{ad}T^l\gamma_4
\ln\frac{4l_c^3}{\pi m_D^2|\nu|}+\cdots,
\eeq
which completes the BRST identity (\ref{eq:BRS4}) to the leading
logarithm level.  The other
order of the limits, $\lim_{\nu^\prime\to\nu}\lim_{\vec p\,^\prime\to\vec p}
\Lambda_j^{l(c)}(P^\prime,P)$ is infrared finite.

To summarise, we have shown how the BRST identity works in a
quark-gluon plasma at high density in terms of Feynman diagrams.  The
derivation of the identity is quite general and that it works at high
density is not surprising, but with this presentation we hope that the
mystery that shrouds this topic may be lifted.  With the incorporation
of HDL, the BRST identity is no longer satisfied order by order.  The
payment for using HDL is that orders of perturbation theory become
mixed up.

\section{Color superconductivity}

\begin{figure}[t]
\epsfxsize 12.0cm
\centerline{\epsffile{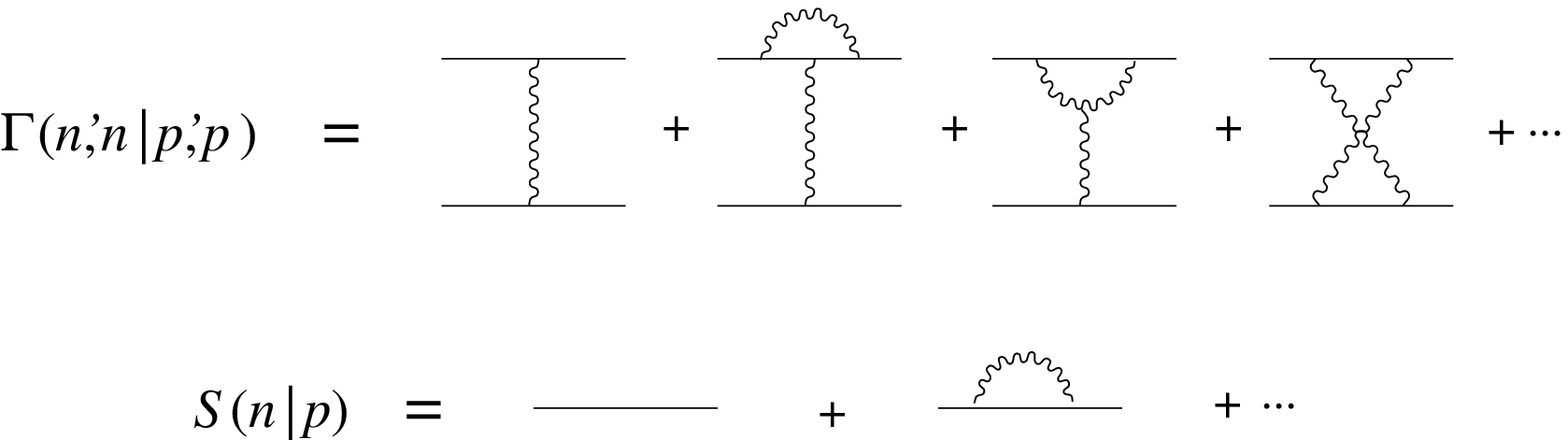}}
\bigskip
%
%
\caption{The diagrammatic expansion of the two particle irreducible vertex
$\Gamma_{s_1^\prime s_2^\prime,
s_1^{\protect\vphantom{\prime}}s_2^{\protect\vphantom{\prime}}}
(n^\prime,n|p^\prime,p)$
to order $g^4$ and the quark self-energy $S_{s's}(n|p)$ to order $g^2$.}
\label{fig5}
\end{figure}

Perturbative QCD has been applied successfully toward the study of color
superconductivity at high baryon densities.  In this regime, single
gluon exchange dominates the pairing interaction, and screening plays an
important role in the non-BCS behavior of color superconductivity
\cite{son1998,schafer1999,pisarski1999b,hong}.  In Refs.~\cite{BLR1,BLR2},
the superconducting pairing temperature of a dense quark-gluon
plasma was investigated by means of a Dyson-Schwinger approach to the
pairing interaction.  The resulting problem was reduced to one of finding the
smallest eigenvalue, $\lambda$, of the Fredholm equation
\beq
f_{s_1^\prime s_2^\prime}(n^\prime|p^\prime)=\frac{\lambda^2}{\beta}
\sum_{n,s_1,s_2}\int_0^\infty dp\,K_{s_1^\prime
s_2^\prime,s_1^{\vphantom{\prime}}s_2^{\vphantom{\prime}}}
(n^\prime,n|p^\prime,p)f_{s_1s_2}(n|p),
\eeq
with the condition $\lambda^2=1$ yielding the critical temperature.
The kernel is given by
\beq
\label{ker}
K_{s_1^\prime s_2^\prime,s_1^{\vphantom{\prime}}s_2^{\vphantom{\prime}}}(n^\prime,n|p^\prime,p)
=\frac{p^2}{2\pi}\sum_{s_1^{\prime\prime}s_2^{\prime\prime}}
\gamma_{s_1^\prime s_2^\prime,s_1^{\prime\prime}s_2^{\prime\prime}}
(n^\prime,n|p^\prime,p)S_{s_1^{\prime\prime}s_1}(n|p)
S_{s_2^{\prime\prime}s_2}(-n|p),
\eeq
and consists of the $s$-wave components of the two particle irreducible
amplitude for the scattering of two quarks in their color antisymmetric
channel with zero
total energy and momentum, $\gamma_{s_1^\prime
s_2^\prime,s_1^{\vphantom{\prime}}s_2^{\vphantom{\prime}}}
(n^\prime,n|p^\prime,p)$, and the full quark propagator $S_{s^\prime s}(n|p)$.
The initial energies of the two quarks are $\pm i\nu_n$ and the final ones are
$\pm i\nu_{n^\prime}$ with $\nu_n=(n+\frac{1}{2})\epsilon$. The initial
momenta of the two quarks are $\pm\vec p$ and the final ones are
$\pm\vec p\,^\prime$. The diagrammatic expansion of
$\gamma_{s_1^\prime s_2^\prime,s_1^{\vphantom{\prime}}s_2^{\vphantom{\prime}}}
(n^\prime,n|p^\prime,p)$ to order $g^4$ and $S_{s's}(n|p)$ to order $g^2$
is shown in Fig.~\ref{fig5}.

Collecting previous results, the perturbative expansion of the least
eigenvalue reads \cite{BLR1,BLR2}
\beqa
\frac{1}{\lambda^2} &=&
{g^2\over24\pi^2}\left(1+{1\over N_c}\right)\left[
{4\over\pi^2}\log^2{2\over\hat\epsilon}
+{8\over\pi^2}(\gamma+\log2)\log{2\over\hat\epsilon}+{\cal O}(1)\right]
\nonumber \\
&&- \left({g^2\over24\pi^2}\right)^2\left(1+{1\over N_c}\right)\left[
C_f {4(\pi^2+4)\over\pi^4}\log^3{2\over\hat\epsilon}
+{\cal O} (\log^2{2\over\hat\epsilon})\right]+{\cal O}(g^6),
\label{eq:plam}
\eeqa
where the leading ${\cal O}(g^2)$ term stems from the first
diagram of Fig.~\ref{fig5} with a bare
quark propagator. Relative to this leading term, the radiative corrections and
the two gluon exchange appear to be suppressed by $g^2$ as is the case with
the remaining diagrams of Fig.~\ref{fig5}, but may not be so
because of the infra-red logarithm, each counted as $g^{-1}$ for $T\sim T_C$ in
accordance with (\ref{eq:scl}). The radiative correction to the quark
propagator is such
an example \cite{son1998}. The logarithm of the self-energy, contained
in the second line of (\ref{eq:plam}), gives rise to a
significant contribution to the prefactor \cite{BLR1}. Though the radiative
correction to the vertex function is liable to such a logarithm according to
the BRST identity, this does not happen in the energy momentum region
$|p-\mu|\sim|\nu|$, $|p^\prime-\mu|\sim|\nu^\prime|$ and
$|\vec p-\vec p\,^\prime|\sim(\kappa|\nu-\nu^\prime|)^{\frac{1}{3}}$, where the
main contribution to the kernel (\ref{ker}) comes from; this is indicated by
the absence of the logarithm in the limit $\nu^\prime\to\nu$ followed by
$\vec p\,^\prime\to\vec p$ of the vertex function. In what follows, we shall
demonstrate this point via an explicit evaluation of the contribution of the
abelian vertex function to the partial wave amplitude.

Consider the abelian vertex Fig.~\ref{fig2}a, with $p=p^\prime=\mu$,
$q=2\mu\sin \frac{\theta}{2}\simeq\mu\theta$. The infra-red contribution
is given by
\beqa
&&\overline\Lambda_j^{(a)}(P',P) = \overline{u}(P')\Lambda_j^{(a)}(P',P)u(P)
\nonumber\\
&&\kern2em
=g^2\hat p_j\int_{l<l_c}\frac{d^3\vec l}{(2\pi)^3}
\int_{-\omega_c}^{\omega_c}\frac{d\omega}{2\pi}{\cal D}(l,\omega)
\frac{\hat p\cdot\hat p^\prime-(\hat p\cdot\hat l)(\hat p^\prime\cdot\hat l)}
{[i(\omega+\nu^\prime)-\xi^\prime][i(\omega+\nu)-\xi]} \\ \nonumber
&&\kern2em
=g^2\hat p_j\int_{l<l_c}\frac{d^3\vec l}{(2\pi)^3}\int_{-\omega_c}^{\omega_c}
\frac{d\omega}{2\pi}{\cal D}(l,\omega)\frac{\hat p\cdot\hat p^\prime
-(\hat p\cdot\hat l)(\hat p^\prime\cdot\hat l)}{i\Delta\nu-\xi^\prime+\xi}
\left[\frac{1}{i(\omega+\nu)-\xi}-\frac{1}{i(\omega+\nu^\prime)-\xi^\prime}
\right],
\eeqa
where $\Delta\nu=\nu^\prime-\nu$, $\xi=|\vec p+\vec l\,|-\mu$ and
$\xi^\prime=|\vec p\,^\prime+\vec l\,|-\mu$ with $|\xi|\leq l$ and
$|\xi^\prime| \leq l$. It follows from the discussions of the previous
sections that the sensitive region of the integration variables which is
responsible to the non-Fermi liquid logarithm corresponds to the
singularities of the fractions inside the bracket. Therefore one of $\xi$
and $\xi^\prime$ must be kept small in the sensitive region. If there were
an infra-red logarithm, it would come from
\beq
\overline\Lambda_\eta^{(a)}(P^\prime,P)=\int_{l<l_c}\frac{d^3\vec l}{(2\pi)^3}
\int_{-\omega_c}^{\omega_c}\frac{d\omega}{2\pi}\frac{{\cal D}(l,\omega)}
{i\Delta\nu-\xi^\prime+\xi}\left[\frac{\theta(\eta l-|\xi|)}{i(\omega+\nu)-\xi}
-\frac{\theta(\eta l-|\xi^\prime|)}{i(\omega+\nu)-\xi^\prime}\right],
\eeq
with $\eta\ll 1$. Transforming the integration variables from $\vec l$ to $l$,
$\xi$ and $\xi^\prime$, we have
\beq
d^3\vec l=\frac{\mu^2}{J}l\,dl\,d\xi\,d\xi^\prime,
\eeq
where the Jacobian is $J=|\vec l\cdot\vec p\times\vec p\,^\prime|\simeq\mu^2
\sqrt{l^2\theta^2-(\xi-\xi^\prime)^2}$ with the approximation that $\xi$ or
$\xi^\prime\ll l$. Introducing
\beq
\overline{\cal D}(l,\omega)=\int_{-\infty}^\omega d\omega^\prime
{\cal D}(l,\omega^\prime),
\eeq
and carrying out the integration over $\xi$ and $\xi^\prime$, we obtain
\beqa
\overline\Lambda_\eta^{(a)}(P^\prime,P) &=& \frac{1}{8\pi^2}\int_0^{l_c}dl
\frac{l}{\sqrt{l^2\theta^2+\Delta\nu^2}}\int_{-\omega_c}^{\omega_c}d\omega
\overline{\cal D}(l,\omega)[\delta(\omega+\nu)-\delta(\omega+\nu^\prime)]
\nonumber \\
&=& \frac{1}{8\pi^2}\int_0^{l_c}dl\frac{l}{\sqrt{l^2\theta^2+\Delta\nu^2}}
\int_{-\nu^\prime}^{-\nu}d\omega {\cal D}(l,\omega).
\eeqa
Note that if $\vec p\,^\prime\to\vec p$ first, we have a complete exposure of
$\Delta\nu$ in the denominator,
\beq
\overline\Lambda_\eta^{(a)}(P^\prime,P)\simeq\frac{1}{8\pi^2}
\frac{1}{\Delta\nu}
\int_0^{l_c}dl\,l\int_{-\nu^\prime}^{-\nu}d\omega {\cal D}(l,\omega).
\eeq
The integration will give rise to $\nu^\prime\ln|\nu^\prime|-\nu\ln|\nu|$,
which in the limit $\nu^\prime\to\nu$ produces the infra-red logarithm. But
here, with $\theta\sim|\nu^\prime-\nu|^{\frac{1}{3}}$ and
$l\sim|\nu|^{\frac{1}{3}}$, such a singularity is suppressed through the
$l^2\theta^2$ term inside of the square root. Indeed, if we insert
$\Lambda_\eta^{(a)}(P^\prime,P)$ into the partial wave integration, we find
the corresponding contribution
\beq
\gamma_{IR}(\nu^\prime,\nu)=\frac{g^4}{8\pi^2}\int_0^{\theta_c}d\theta\,\theta
{\cal D}(\mu\theta,\nu^\prime-\nu)
\int_0^{l_c}dl\frac{l}{\sqrt{l^2\theta^2+\Delta\nu^2}}
\int_{-\nu^\prime}^{-\nu}d\omega {\cal D}(l,\omega).
\eeq
$\gamma_{IR}(\nu^\prime,\nu)$ may be bounded by dropping $\Delta\nu$ inside
the square root. Then the integration over $\theta$ decouples
from that over $l$ and $\omega$, {\it i.e.}
\beq
|\gamma_{IR}(\nu^\prime,\nu)|\leq \frac{g^4}{8\pi^2}IJ,
\eeq
where
\beq
I=\int_0^{\theta_c}d\theta {\cal D}(\mu\theta,\nu^\prime-\nu)
\eeq
and
\beq
J=\int_0^{l_c}dl\int_{-\nu'}^{-\nu}d\omega {\cal D}(l,\omega).
\eeq
It follows from the properties of the function ${\cal D}(l,\omega)$ that
$I$ and $J$ are bounded from above by
\beq
I\leq \frac{2\pi}{3\sqrt{3}\mu}(\kappa|\nu^\prime-\nu|)^{-\frac{1}{3}}
\eeq
and
\beq
J\leq \frac{\pi}{\sqrt{3}}\kappa^{-\frac{1}{3}}||\nu^\prime|^{\frac{2}{3}}
-{\rm{Sign}}(\nu^\prime\nu)|\nu|^{\frac{2}{3}}|,
\eeq
where $\kappa={\pi\over4}m_D^2$.
Combining $I$ and $J$, we see that $\gamma_{IR}(\nu^\prime,\nu)$ is
nonsingular in the limit $\nu^\prime\to 0$ and $\nu\to 0$ along any
path in the $(\nu^\prime,\nu)$-plane.

It is important to note that this result for $\gamma_{IR}(\nu^\prime,\nu)$
only pertains to any possible additional infra-red enhancement arising from
the radiative quark-gluon vertex $\Lambda^{l(a)}_\eta$.  For the complete
partial wave amplitude, $\gamma(\nu',\nu)$, the collinear magnetic gluon
exchange logarithm, already present at tree level, {\it i.e.}
\begin{equation}
\gamma_{\rm tree}(\nu',\nu)\simeq
{g^2\over6\mu^2}\ln{8\mu^3\over\kappa|\Delta\nu|},
\end{equation}
maintains its presence at the radiative level.  Indeed, based on
numerical evaluation of the partial wave amplitude,
$\gamma_{\rm abelian}(\nu^\prime, \nu)$, we have confirmed that only this
expected collinear logarithm is present.  We have also evaluated the
infra-red contribution $\gamma_{IR}(\nu^\prime, \nu)$ numerically, and
the result supports the above analytic arguments.

Though our conclusion that the vertex function does not contribute to the
pre-exponential factor agrees with that made in \cite{schafer1999},
the arguments used in \cite{schafer1999} to justify this conclusion
merit further consideration.  In particular, the formula for
the vertex function, taken from Ref.~\cite{Bellac}, is not applicable
for the infra-red contribution at a large chemical potential in
comparison with the temperature.
This can be judged by the absence of the infra-red logarithm from their
vertex function in any order of the limit of zero energy-momentum transfer;
this absence contradicts the BRST identity as discussed above. In fact,
only the expressions for diagrams with internal fermion lines only can be
carried over from the high temperature region to the large chemical
potential region, as is the case with the gluon self-energy functions
(\ref{eq:F}) and (\ref{eq:G}). For diagrams with internal gluon lines,
the infra-red region makes significant contributions, leading to effects
such as the non-Fermi liquid behaviour of the quark self-energy and
vertex functions,
which has been completely ignored by the Hard thermal loop approximation
employed in \cite{Bellac}.

By careful examination of the radiative corrections to the quark
self-energy and vertex functions, we have reconciled the non-Fermi
liquid behavior in the dense plasma with the BRST identity.  The
incorporation of HDL, and the resulting resummation in the gluon
propagator, leads to a mixing of orders in the perturbative expansion.
Hence proof of BRST involves combining diagrams of different loop order,
as seen in Fig.~\ref{fig2}.  An important consequence of this result for
color superconductivity is the verification that there are no additional
infra-red logarithms accompanying the radiative correction to the vertex
function.  This strengthens our previous result that the only radiative
correction to the determination of $T_C$ comes from the quark
self-energy, and suggests that the pre-exponential factor of
(\ref{eq:ratio}) is in fact exact to leading order in $g$.

\bigskip
\noindent
{\bf Acknowledgements.}

We would like to thank R.~Pisarski and D.~Rischke for raising the
issue of the consistency of the non-Fermi liquid behavior with the
BRST identity, which motivated this investigation.  We also wish to
thank D.T.~Son for bringing Ref.~\cite{Chakra} to our attention.
The work of W.E.~Brown and H.C.~Ren is supported in part by the US
Department of Energy under grant DOE-91ER40651-TASKB.
H.C.~Ren's work is also supported in part by the Wiessman visiting
professorship of Baruch College of CUNY.

Hai-cang Ren would like to dedicate this work to his friend,
Dr.~D.Y.~Chen, who passed away following a tragic accident.

\begin{appendix}

\section{}

In this appendix, we shall prove the BRST identity, (\ref{BRST}), relating
the self-energy, vertex and ghost diagrams of Figs.~\ref{fig1}, \ref{fig2}
and \ref{fig3} in the presence of hard dense loops.

Using the standard trick,
\beq
(P' - P)^{\mu}S( P' + L) \gamma_{\mu} S(P + L) = S( P' + L)
- S( P +L),
\eeq
we may trivially relate the abelian vertex Fig.~\ref{fig2}a with self-energy
Fig.~\ref{fig1}.
\beq
\label{res1}
(P' -P)^{\mu} \Lambda_{\mu}^{l(a)}(P',P) = gT_f^l \left(1 - \frac{C_{ad}}{2C_f}
\right)\left( \Sigma(P') - \Sigma(P) \right).
\eeq
which, apart from the group theoretic factors, is nothing but the Takahashi
identity of QED and is independent of the form of the gluon propagator.
However, for non-abelian gauge theories, the group coefficients do not
match; cancellation of the extra term must result from the additional
vertices.

In QCD there is a second physical process at $O(g^3)$ in perturbation
theory, formed with the tri-gluon vertex $-if^{lmn}\Gamma_{\mu \lambda
\rho}$, as shown in Fig.~\ref{fig2}b.  We now turn to this diagram
to see how it may cancel the extra terms. It is straightforward to write
down an expression,
\beq
\label{trigluon}
\Lambda_{\mu}^{l(b)}(P',P) = f^{lmn} T_f^m T_f^n \frac{g^3}{\beta} \sum_n
\int \frac{d^3 \vec{l}}{(2\pi)^3} {\mathcal D}_{\nu \lambda}(L)
(-i)\Gamma_{\mu \lambda \rho}(L,L-Q) {\mathcal D}_{\rho \nu'}(L)
\gamma_{\nu} S(P+L) \gamma_{\nu'},
\eeq
where $i f^{lmn} T_f^m T_f^n = -\frac{C_{ad}}{2} T_f^l$.  However,
this expression may be simplified when contracted with $(P^\prime-P)_\mu$
with the aid of the identity
\beq
\label{tri}
(P' - P)^{\mu} {\mathcal D}_{\nu' \lambda}(P') \Gamma_{\mu \lambda
\rho} (P',P) {\mathcal D}_{\rho \nu}(P) = i \left\{
V_{\nu'\nu}^{(1)}(P',P) +
V_{\nu'\nu}^{(2)}(P',P) + V_{\nu'\nu}^{(3)}(P',P) \right\},
\eeq
where
\beqa
\label{eq:tri3}
V_{\nu'\nu}^{(1)}(P',P) &=& i \left[
{\mathcal D}_{\nu' \nu}(P) - {\mathcal D}_{\nu' \nu}(P')\right], \\ \nonumber
V_{\nu'\nu}^{(2)}(P',P) &=&  {\mathcal D}_{\nu' \lambda}(P')
\left[\Pi_{\lambda \rho}
(P') - \Pi_{\lambda \rho}(P) \right] {\mathcal D}_{\rho \nu}(P), \\ \nonumber
V_{\nu'\nu}^{(3)}(P',P) &=&  \Delta(P') P_{\nu'}'
P_{\lambda}^\prime {\mathcal D}_{\lambda \nu}(P)-
{\mathcal D}_{\nu' \lambda}(P') P_{\lambda} P_{\nu} \Delta(P) .
\eeqa
$\Pi_{\mu \nu}(P)$ is the HDL diagram which satisfies
$P_\mu\Pi_{\mu\nu}(P)=0$ and $\Delta(P)=-i/p^2$ is
the ghost propagator.
Since $\Pi$ is itself of $O(g^2)$, we see here that the price one pays for
incorporating HDL in the gluon propagator is the mixing of orders in
perturbation theory.  To prove (\ref{tri}), we start with the bare gluon
propagator,
\beq
\label{eq:bgp}
D_{\mu\nu}(P)=\frac{-i}{P^2}\Big[\delta_{\mu\nu}+(\alpha-1)
\frac{P_\mu P_\nu}{P^2}\Big],
\eeq
and the identity
\beq
\label{eq:amatrix}
(P^\prime-P)_\mu(-i)\Gamma_{\mu\rho\lambda}(P^\prime,P)=
(P^2-P^{\prime 2})\delta_{\rho\lambda}+P_\rho^\prime P_\lambda^\prime
-P_\rho P_\lambda.
\eeq
Sandwiching (\ref{eq:amatrix}) between $D(P^\prime)$ and $D(P)$, we find
\begin{eqnarray}
\label{eq:dad}
-i(P^\prime-P)_\mu &&D_{\alpha^\prime\rho}(P^\prime)\Gamma_{\mu\rho\lambda}
(P^\prime,P)D_\alpha(P)\\
&&\quad=-i[D_{\alpha^\prime\alpha}(P^\prime)-D_{\alpha^\prime
\alpha}(P)]+\Delta(P^\prime)P_{\alpha^\prime}^\prime P_\rho^\prime
D_{\rho\alpha}(P)-\Delta(P)D_{\alpha\rho}(P^\prime)P_\rho P_\alpha.
\nonumber
\end{eqnarray}
The HDL-dressed gluon propagator
is related to the bare propagator via the Dyson-Schwinger equation,
\beq
\label{eq:DS}
{\cal D}_{\mu\nu}(P)=D_{\mu\nu}(P)-iD_{\mu\rho}(P)\Pi_{\rho\lambda}(P)
{\cal D}_{\lambda\nu}(P)=D_{\mu\nu}(P)-i{\cal D}_{\mu\rho}(P)
\Pi_{\rho\lambda}(P)D_{\lambda\nu}(P).
\eeq
It follows that $-i(P^\prime-P)_\mu{\cal D}_{\nu^\prime\lambda}(P^\prime)
\Gamma_{\mu\lambda\rho}(P^\prime,P){\cal D}_{\rho\nu}(P)$ can be obtained
by sandwiching (\ref{eq:dad}) between
$\delta_{\nu^\prime\alpha^\prime}-i{\cal D}_{\nu^\prime\beta^\prime}(P^\prime)
\Pi_{\beta^\prime\alpha^\prime}(P^\prime)$ on the left and
$\delta_{\alpha\nu}-i\Pi_{\alpha\beta}(P){\cal D}_{\beta\nu}(P)$
on the right. The expression is then simplified by the 4-dimensional
transversality of the self-energy matrix $\Pi(p)$, and we end up with
(\ref{tri}) and (\ref{eq:tri3}).

Now we look at the contribution due
to $V^{(1)}$,
\beq
i\frac{C_{ad}}{2} T_f^l \frac{g^3}{\beta} \sum_n \int \frac{d^3\vec{l}}{
(2\pi)^3} \left[ {\mathcal D}_{\nu' \nu}(L-Q) - {\mathcal D}_{\nu' \nu}(L)
\right] \gamma_{\nu} S(P+L) \gamma_{\nu'} = ig\frac{C_{ad}}{2 C_f} T_f^l
\left(\Sigma(P') - \Sigma(P) \right).
\eeq
This expression, the origins of which are purely non-abelian in nature,
exactly cancels the extra terms induced in (\ref{res1}).  However, the
tri-gluon vertex also induces a number of extra terms which we shall
now consider in turn.

The appearance of the ghost propagators in
$V^{(3)}$ suggests that these extra terms will be cancelled by
the non-physical ghost-quark vertices generated by the BRST
transformations.  Indeed, we find that, when grown from a fermion line,
the ghost terms contribute,
\beq
if^{lmn} T_f^m T_f^n \frac{g^3}{\beta} \sum_n \int \frac{d^3\vec{l}}{
(2\pi)^3}  V_{\nu'\nu}^{(3)}(L,L-Q)\gamma_{\nu'} S(L+P) \gamma_{\nu},
\eeq
which exactly cancels the diagrams of Fig.~\ref{fig3}.

The remaining term, $V^{(2)}$, is of $O(g^5)$, two orders higher
in perturbation theory, and contributes,
\beq
\label{remainder}
i f^{lmn}T_f^m T_f^n \frac{g^3}{\beta} \sum_n \int\frac{d^3\vec{l}}{
(2\pi)^3}  {\mathcal D}_{\nu' \lambda} (L-Q)
\left[\Pi_{\lambda \rho}
(L) - \Pi_{\lambda \rho}(L-Q) \right]
{\mathcal D}_{\rho \nu}(L) \gamma_{\nu'} S(L+P) \gamma_{\nu}.
\eeq
In the absence of HDL this term does not appear and the BRST identity
is satisfied order by order in perturbation theory.  Although with the
inclusion of HDL the ordering has become mixed up, the identity must
remain.  To see how this contribution is cancelled we study the
triangular vertex shown in Fig.~\ref{fig2}c.
\begin{figure}[t]
\epsfxsize 1.5cm
\centerline{\epsffile{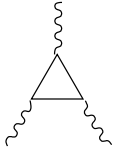}}
\bigskip
\caption{Quark loop with three external gluons,
$\widetilde{\Gamma}_{\mu \lambda \rho}^{lmn}$.}
\label{fig4}
\end{figure}

We shall first look at one of the two loops that form the triangular
vertex, namely the quark loop with three external gluons, as shown in
Fig.~\ref{fig4}.  With three identical vertices there are two possible
orderings for this diagram.  Considering both orderings, we write this
vertex correction as
\beqa
\label{eq:vc}
\widetilde{\Gamma}_{\mu\lambda\rho}^{lmn} (P',P) = \frac{g^3}{\beta} \sum_n
\int \frac{d^3 \vec{l}}{(2\pi)^3}&&{\rm Tr} \left[ T_f^l T_f^n T_f^m
\gamma_{\lambda} S(L + P - P') \gamma_{\mu} S(L) \gamma_{\rho} S(L+P) \right.
\\ \nonumber
&& + \left.
T_f^l T_f^m T_f^n \gamma_{\lambda} S(-L-P) \gamma_{\rho} S(-L) \gamma_{\mu}
S(-L-P+P') \right].
\eeqa
Contracting one leg of the triangular vertex with $(P' - P)^{\mu}$ we find,
\beq
\label{eq:clv}
(P'-P)^{\mu}\widetilde{\Gamma}_{\mu \lambda \rho}^{lmn} (P',P) =
i g f^{lmn} \left[ \Pi_{\lambda \rho}(P') - \Pi_{\lambda \rho}(P) \right],
\eeq
where, as discussed, $\Pi$ is the vacuum polarization diagram.  Therefore,
connecting the two free legs to a fermion line, we find that
\beqa
& & (P' -P)^{\mu} \Lambda_{\mu}^{l(c)}(P',P) = \nonumber \\
& &  - f^{lmn} T_f^m T_f^n \frac{g^3}{
\beta} \sum_n \int\frac{d^3 \vec{l}}{(2\pi)^3} {\mathcal D}_{\nu' \lambda}(L-Q)
\left[ \Pi_{\lambda \rho}(L-Q) - \Pi_{\lambda \rho}(L) \right] {\mathcal D}_{
\rho\nu}(L) \gamma_{\nu'} S(L+P) \gamma_{\nu}, \nonumber \\
\eeqa
which cancels the remainder in (\ref{remainder}).

At this point, we note that at nonzero chemical potential the triangular
diagram, (\ref{eq:vc}), does not contain exclusively the term proportional
to $f^{lmn}$, and is nonvanishing even in QED because of the breakdown of
Furry's theorem by the Fermi sea. On the other hand, the identity
(\ref{eq:clv}) remains
valid rigorously and there is no contribution
from the triangular diagram to the Takahashi identity of QED. Furthermore, for
low excitations near the Fermi level, the approximate particle-hole symmetry
renders the triangular diagram dominated by the term proportional to
$f^{lmn}$.

Before concluding this appendix, we shall relate the particular BRST
identity (\ref{BRST}) to the master BRST identity as given in
Ref.~\cite{itzykson}.  Let $\Gamma(A,\chi,\bar\chi, c,\bar c)$ be
the generating functional of proper vertex functions with
$A$, $\chi$, $\bar\chi$, $c$ and $\bar c$ the quantum mechanical average of the
gauge potential, $V_\mu$, quark fields $\psi$, $\bar\psi$ and the ghost fields
$\phi$, $\bar\phi$, {\it i.e.}, $A_\mu(x)=\langle V_\mu(x)\rangle$,
$\chi(x)=\langle\psi(x)\rangle$, $\bar\chi(x)=\langle\bar\psi(x)\rangle$,
$c(x)=\langle\phi(x)\rangle$ and $\bar c(x)=\langle\bar\phi(x)\rangle$.
The master BRST identity reads
\beq
\label{eq:mbrst}
\int d^4x\Big[\frac{\delta \Gamma}{\delta A_\mu(x)}\langle\delta
V_\mu(x)\rangle
+\frac{\delta \Gamma}{\delta \chi(x)}\langle\delta\psi(x)\rangle
+\frac{\delta \Gamma}{\delta \bar\chi(x)}\langle\delta\bar\psi(x)\rangle
+\frac{\delta \Gamma}{\delta c(x)}\langle\delta\phi(x)\rangle\Big]=0,
\eeq
where
\begin{eqnarray}
\label{eq:da}
\delta V_\mu^l&=&\frac{\partial \phi^l}{\partial x_\mu}+gf^{lmn}
V_{\mu}^m\phi^n,\\
\label{eq:dpsi}
\delta\psi&=&-iT^l\phi^l\psi,\\
\label{eq:dbar}
\delta\bar\psi&=&-iT^l\phi^l\bar\psi,\\
\delta\phi^l&=&\frac{1}{2}f^{lmn}\phi^m\phi^n,
\end{eqnarray}
are the BRST variations of the field components. The expansion of the term
$\bar\chi\chi c$ in (\ref{eq:mbrst}) to the order $g^3$ and with the
bare gluon propagators replaced by the dressed ones afterwards
yield the identity (\ref{BRST}). Unlike an abelian
gauge theory, the ghosts couple to other fields of the theory. The expectation
of the nonlinear term of the BRST variations gives rise to the additional terms
$R^a(p^\prime,p)$ with $R^{l(a)}$ from the second term of (\ref{eq:da}),
$R^{l(b)}$ from (\ref{eq:dpsi}) and $R^{l(c)}$ from (\ref{eq:dbar}).

\section{}

In this appendix, we shall evaluate the infra-red contribution of the diagram
in Fig.~\ref{fig2}c, which we denote by $\Lambda_\mu^{l(c)}(P^\prime,P)_{IR}$
with $P=(\vec p,\nu)$ and $P^\prime=(\vec p+\vec q,\nu+\Delta\nu)$. Then
$Q=P^\prime-P=(\vec q,\Delta\nu)$, and both $\vec q$ and $\Delta\nu$ are
soft. The calculation is greatly simplified with the aid of the identity
(\ref{eq:clv}) for $\mu=4$ in the limit $\vec q\to 0$ followed by
$\Delta\nu\to 0$ and for $\mu=j$
in the limit $\Delta\nu\to 0$ followed by $\vec q\to 0$.

($i$) The triangular vertex in the limit
\begin{equation}
\lim_{\Delta\nu\to 0}\lim_{\vec q\to 0}
\Lambda_4^{l(c)}(P^\prime,P)_{IR}.
\end{equation}
We start with
\beqa
\label{l4}
\Lambda_4^{l(c)}(P^\prime,P)_{IR}&=&-g^2\int_{l<l_c}\frac{d^3\vec l}{(2\pi)^3}
\int_{-\omega_c}^{\omega_c}\frac{d\omega}{2\pi}T^aT^b[-i\widetilde
\Gamma_{4m^\prime n^\prime}^{lab}(L,L-Q)]
\gamma_m\frac{i}{p+l}\gamma_n \nonumber \\
&\times& {\cal D}(|\vec l-\vec q\,|,\omega-\Delta\nu)
{\cal D}(l,\omega)\left[\delta_{m^\prime m}
-{(\vec l-\vec q\,)_{m^\prime}(\vec l-\vec q\,)_m\over|\vec l-\vec
q\,|^2}\right](\delta_{n^\prime n}-\hat l_{n^\prime}\hat l_n),
\eeqa
where ${\cal D}(l,\omega)$ is given by (\ref{eq:calddef}) and
$\widetilde\Gamma_{\mu\nu\rho}^{lmn}$ by Fig.~\ref{fig4}. Note that we have
used the continuum approximation for the Matsubara sum. It
follows from (\ref{eq:clv}) that
\beq
\lim_{\Delta\nu\to 0}\lim_{\vec q\to 0}\widetilde\Gamma_{4ij}^{lab}(L,L-Q)
=igf^{lab}\frac{\partial}{\partial\omega}\Pi_{ij}(L).
\eeq
Therefore
\beq
\lim_{\Delta\nu\to 0}\lim_{\vec q\to 0}\Lambda_4^{l(c)}(P^\prime,P)_{IR}
=\frac{1}{2}gC_{ad}T^l\Lambda(P),
\eeq
with
\beqa
\Lambda(P) &=& -g^2\int_{l<l_c}\frac{d^3\vec l}{(2\pi)^3}
\int_{-\omega_c}^{\omega_c}d\omega\frac{\partial\sigma^M}{\partial\omega}
{\cal D}^2(l,\omega)
\frac{\gamma_4-i(\hat p\cdot\hat l)^2\gamma\cdot\hat p}{i(\omega+\nu)-\xi}
\\
&=&\frac{ig^2}{8\pi^3}\int_0^{l_c}dl\,l\int_{-\omega_c}^{\omega_c}d\omega
\frac{\partial\sigma^M}{\partial\omega}\frac{1}{[l^2+\omega^2+\sigma^{M}
(l,\omega)]^2}\int_{-l}^ld\xi
(\gamma_4-i\frac{\xi^2}{l^2}\vec\gamma\cdot\hat p)\frac{1}{i(\omega+\nu)-\xi},
\nonumber
\eeqa
where $p=\mu$ and $\xi=|\vec p+\vec l\,|-\mu$. Carrying out the
$\xi$ integration, we find
\beq
\Lambda(P)=\frac{ig^2}{2\pi^3}\int_0^{l_c}\frac{dl}{l}
\int_{-\omega_c}^{\omega_c}
d\omega\frac{\partial\sigma^M}{\partial\omega}\frac{F(\nu,\mu;l,\omega)}
{[l^2+\omega^2+\sigma^{M}(l,\omega)]^2},
\eeq
with $F(\nu,\mu;l,\omega)$ given by (\ref{eq:eff})
The discontinuity of the inverse tangent corresponds to $\omega\sim-\nu$ and
the $l$-integration is dominated at $l\sim (\kappa\omega)^{\frac{1}{3}}\sim
(\kappa\nu)^{\frac{1}{3}}$. We end up with
\beqa
\Lambda(P) &=& -\frac{g^2}{4\pi^3}\int_0^{l_c}\frac{dl}{l}
\int_{-\omega_c}^{\omega_c}
d\omega\frac{\partial\sigma^M}{\partial\omega}\frac{F(\nu,\mu;l,\omega)}
{[l^2+\sigma^{M}(l,\omega)]^2}
\nonumber \\
&=& \frac{g^2}{4\pi^2}\gamma_4\int_0^{l_c}dl{\cal D}(l,-\nu)
+\hbox{terms regular as $\nu\to 0$} \nonumber \\
&=& \frac{ig^2}{12\pi^2}\gamma_4\log\frac{l_c^3}{\kappa|\nu|}
+\hbox{terms regular as $\nu\to 0$}.
\eeqa

($ii$) The triangular vertex in the limit
\begin{equation}
\lim_{\vec q\to 0}\lim_{\Delta\nu\to 0}
\Lambda_j^{l(c)}(P^\prime,P)_{IR}.
\end{equation}
Here $\Lambda_j^{l(c)}(P',P)$ is given by (\ref{l4}) with the
replacement $\widetilde\Gamma_{4m'n'}^{lmn}\to
\widetilde\Gamma_{jm'n'}^{lmn}$.  It then follows from the identity
(\ref{eq:clv}) that
\beq
\lim_{\vec q\to 0}\lim_{\Delta\nu\to 0}\widetilde\Gamma_{jmn}^{lab}(L,L-Q)
=igf^{lab} \frac{\partial}{\partial l_j}\Pi_{mn}(L).
\eeq
Therefore
\beq
\lim_{\vec q\to 0}\lim_{\Delta\nu\to 0}\Lambda_j^{l(c)}(P^\prime,P)_{IR}
=\frac{1}{2}gC_{ad}T^l\hat p_j\Lambda^\prime(P),
\eeq
with
\beqa
\Lambda^\prime(P) &=& -\frac{g^2}{8\pi^3}\int_0^{l_c}dl
\int_{-\omega_c}^{\omega_c}d\omega
\frac{\partial\sigma^M}{\partial l}\frac{1}{[l^2+\omega^2+\sigma^{M}
(l,\omega)]^2}\int_{-l}^ld\xi\,\xi
(\gamma_4-i\frac{\xi^2}{l^2}\vec\gamma\cdot\hat p)\frac{1}{i(\omega+\nu)-\xi}
\nonumber \\
&=& -\frac{g^2}{4\pi^3}m_D^2\int_0^{l_c}dl\int_{-\omega_c}^{\omega_c}d\omega
\frac{\partial f^M}{\partial l}\frac{l}
{\left[l^2+\omega^2+m_D^2f^M({\omega\over l})\right]^2}\nonumber\\
&&\kern8em\times\Biggl\{
\gamma_4\left(-1+{\omega+\nu\over l}\tan^{-1}{l\over\omega+\nu}\right)
\nonumber\\
&&\kern9.5em
+i\left[\left({1\over3}-{(\omega+\nu)^2\over l^2}\right)
+{(\omega+\nu)^3\over l^3}\tan^{-1}{l\over\omega+\nu}\right]\vec\gamma
\cdot\hat p\Biggr\}.
\eeqa
The discontinuity of the inverse tangent function at $\omega+\nu=0$ is now
smeared by the factor $\omega+\nu$.  This integral converges in the
limit $\nu\to0$.

\end{appendix}



\begin{references}

\bibitem{bailin1984} D. Bailin and A. Love, {\sl Superfluidity and
Superconductivity in Relativistic Fermion Systems},
Phys. Rep. {\bf 107}, 325 (1984), and references therein.

\bibitem{alford1998a} M. Alford, K. Rajagopal and F. Wilczek, {\sl QCD
at Finite Baryon Density: Nucleon Droplets and Color
Superconductivity}, Phys. Lett. {\bf B422}, 247 (1998).

\bibitem{rapp1998} R. Rapp, T. Schaeffer, E.V. Shuryak and
M. Velkovsky, {\sl Diquark Bose Condensates in High Density Matter and
Instantons}, Phys. Rev. Lett. {\bf 81}, 53 (1998).

\bibitem{alford1999} M. Alford, K. Rajagopal and F. Wilczek,
{\sl Color-Flavor Locking and Chiral Symmetry Breaking in High Density QCD},
Nucl. Phys {\bf B537}, 443 (1999).

\bibitem{schafer1998} T. Sch\"afer and F. Wilczek, {\sl Continuity of
Quark and Hadron Matter}, Phys. Rev. Lett. {\bf 82}, 3956 (1999).

\bibitem{pr1} R.D. Pisarski, D.H. Rischke,
{\sl A First Order Transition and Parity Violation in a Color Superconductor},
Phys. Rev. Lett {\bf 83}, 37 (1999).

\bibitem{pr2} R.D. Pisarski, D.H. Rischke,
{\sl Superfluidity in a model of massless fermions coupled to scalar bosons},
Phys.  Rev. {\bf D60}, 094013 (1999).

\bibitem{hdl1}
J.P. Blaizot and J.Y. Ollitrault,
{\sl Collective fermionic excitations in systems with a large chemical
potential}, Phys. Rev. {\bf D48}, 1390 (1993);

\bibitem{hdl2}
H. Vija and M.H. Thoma,
{\sl Braaten-Pisarski Method at Finite Chemical Potential},
Phys. Lett. {\bf B342}, 212 (1995).

\bibitem{hdl3}
C. Manuel, {\sl Hard dense loops in a cold non-Abelian plasma},
Phys. Rev. {\bf D53}, 5866 (1996).

\bibitem{Gell-Mann} M. Gell-Mann and K. A. Brueckner, Phys. Rev. {\bf 106},
364 (1957).

\bibitem{son1998} D.T. Son, {\sl Superconductivity by long-range color
magnetic interaction in high-density quark matter},
Phys. Rev. {\bf D59}, 094019 (1999).

\bibitem{pincus} T. Holstein, R.E. Norton and P. Pincus, Phys. Rev. {\bf B6},
2649 (1973); M. Yu. Reizer, Phys. Rev. {\bf B40}, 11571 (1988), Phys. Rev.
{\bf B44}, 5476 (1991).

\bibitem{varma} C.M. Varma, P.B. Littlewood, S. Schmitt-Rink, E. Abrahams
and A. E. Ruckenstein, Phys. Rev. Lett. {\bf 66}, 1996 (1989).

\bibitem{schafer1999} T. Sch\"afer and F. Wilczek, {\sl Superconductivity
from perturbative one-gluon exchange in high density quark matter},
Phys. Rev. {\bf D60}, 114033 (1999).

\bibitem{pisarski1999b} R.D. Pisarski, D.H. Rischke, {\sl Gaps and critical
temperature for color superconductivity}, Phys. Rev. {\bf D61}, 051501
(2000).

\bibitem{hong} D.K. Hong, V.A. Miransky, I.A. Shovkovy
and L.C.R. Wijewardhana, {\sl Schwinger-Dyson approach to color
superconductivity in dense QCD}, Phys. Rev. {\bf D61}, 056001 (2000).

\bibitem{BLR1} W.E. Brown, J.T. Liu and H.C. Ren,
{\sl On the Perturbative Nature of Color Superconductivity},
{\tt hep-ph/9908248}, to appear in Phys. Rev. D.

\bibitem{pisarski1999c} R.D. Pisarski, D.H. Rischke,
{\sl Color superconductivity in weak coupling}, Phys. Rev. {\bf D61},
074017 (2000).

\bibitem{BLR2} W.E. Brown, J.T. Liu and H.C. Ren,
{\sl The Transition Temperature to the Superconducting Phase of QCD at
High Baryon Density}, {\tt hep-ph/9912409}.

\bibitem{Bellac} M. Le Bellac, {\it Thermal Field Theory},
Cambridge University Press (1996).

\bibitem{damp1}
M. Le Bellac and C. Manuel,
{\sl Damping rate of quasiparticles in degenerate ultrarelativistic plasmas},
Phys. Rev. {\bf D55}, 3215 (1997).

\bibitem{damp2}
B. Vanderheyden and J.Y. Ollitrault, 
{\sl Damping rates of hard momentum particles in a cold
ultrarelativistic plasma}, Phys. Rev. {\bf D56}, 5108 (1997).

\bibitem{Chakra} S. Chakravarty, R.E. Norton and O.F. Syljuasen,
{\sl Transverse gauge interactions and the vanquished Fermi liquid},
Phys. Rev. Lett. {\bf 74}, 1423 (1995).

\bibitem{itzykson} C. Itzykson and J. Zuber, {\it{Quantum Field Theory}},
McGraw-Hill (1980).

\end{references}
\end{document}